\documentclass[numerical,amssymb,aip,jmp,preprint]{revtex4-1}

\usepackage{graphicx}%
\usepackage{dcolumn}%
\usepackage{bm}%

\usepackage{amsthm} 
\usepackage{amsmath}
\usepackage{amstext}
\usepackage{amsfonts}
\usepackage{amssymb,amscd}
\usepackage{amsopn}
\usepackage{graphicx}
\usepackage{xcolor}

\newcommand{\etal}{\textit{et al.}}
\newtheorem{twr}{Theorem} 
\newtheorem{property}[twr]{Property} 
\newtheorem{lem}[twr]{Lemma} 
 
\newtheorem{OurCorollary}[twr]{Corollary} 
\newtheorem{obs}[twr]{Observation} 
 
\newtheorem{den}{Denotation} 

\usepackage[utf8]{inputenc}  
\usepackage[T1]{fontenc} 

\begin{document}

\title{Dynamical semigroups in the Birkhoff polytope of order 3 as a tool for analysis of quantum channels}

\author{Mateusz Snamina}
\email{mateusz.snamina@doctoral.uj.edu.pl}
\affiliation{Department of Theoretical Chemistry, Jagiellonian University, ul. Ingardena 3, 30-060 Krakow, Poland}

\author{Emil J. Zak}
\email{emil.j.zak@gmail.com	}
\affiliation{Chemistry Department, Queen's University, 90 Bader Lane, K7L 3N6 Kingston, ON, Canada}
%\affiliation{Department of Physics and Astronomy, University College London, Gower Street, London WC1E 6BT, United Kingdom}
\date{\today}

\begin{abstract}
In the present paper we show a link between bistochastic quantum channels and classical maps.
The primary goal of this work is to analyse the multiplicative structure of the Birkhoff polytope of order 3 (the simplest non-trivial case).
A suitable complex parametrization of the Birkhoff polytope is proposed, which reveals several its symmetries and characteristics, in particular: (i) the structure of Markov semigroups inside the Birkhoff polytope,
(ii) the relation between the set of Markov time evolutions, the set of positive definite matrices and the set of divisible matrices. A condition for Markov time evolution of semigroups in the set of symmetric bistochastic matrices is then derived, which leads to an universal conserved quantity for all Markov evolutions.
Finally, the complex parametrization is extended to the Birkhoff polytope of order 4.
\end{abstract}

%\begin{keywords}
%  Continuous time Markov system, Stochastic systems;
%\end{keywords}

\maketitle

\section{Introduction}

Recent experimental and theoretical developments in the field of quantum information theory attracted a considerable attention to the dynamics of quantum open systems.
The density operator, as a central object in the quantum theory of open systems undergoes time evolution, which is in general non-unitary, due to interaction with the environment \cite{GeomOfQState,TheTheoryOfOpenQuantumSystems,Nielsen}.
Instead, completely positive (CP) trace preserving (TP) linear maps need to be introduced, which transform the initial state of the system (initial density operator) into an arbitrary time-advanced state; these maps are identified with quantum channels, which remain of great importance both for the physics of quantum open systems and the information theory. A comprehensive review on CPTP maps was given by Kraus \cite{Kraus}.

Time evolution of a quantum system may be analysed at different levels of approximation.
By considering time independent Hamiltonians, as well as neglecting memory effects (mediated by the environment),
one obtains the \textit{homogeneous Markov evolution} \cite{EvolOfAttainableStructuresOfHMS,HMS_AsAnElasticMedium,Rivas2012}.
The corresponding dynamical semigroup of quantum channels is then described by the Kossakowski -- Lindblad master equation \cite{kossakowski1976,Lindblad1976,Havel}.
In a general case, memory effects are present, imposing the necessity for a more complicated approach based on the Nakijama-Zwanzig equation \cite{Rivas2012,chruscinski2012,chruscinski2014}. Despite the existence of general formulas, the problem of finding, among all possible channels, these realizing Markov evolution still brings much effort \cite{Rivas2012,Wolf2007,Aniello2013}. This issue has been analysed by \textit{Wolf, et al.} in Ref.~\cite{DividingQuCh}, where a classification of quantum channels is introduced (in particular with respect to their ability to represent a simple Markov dynamics).

The intrinsic structure of the set of quantum channels is non-trivial even in the simplest case of the evolution of a qubit.
Despite numerous classical approaches \cite{HMS_AsAnElasticMedium,A_structure_of_doubly_stochastic_Markov_chains}, to the best of our knowledge, the problem of classification of quantum channels (in terms introduced in Ref.~\cite{DividingQuCh})
defined in a three-dimensional Hilbert space remains open.
Therefore, the present paper aims in a preliminary analysis and classification of quantum channels in the 3D Hilbert space.

A considerable complexity of the problem creates a demand for selection of a subset of quantum channels which exhibit physical significance.
An important example is a class of channels for which the steady-state corresponds to the maximal entropy state. Such channels are called \textit{unital} \cite{Wolf2009,Evans1984}.
In the 2D case the unital channels are often named \textit{Pauli channels}, which contain operations such as: bit flip, phase flip or depolarizing channel \cite{Nielsen}. For a comprehensive review see Refs. \cite{GeomOfQState,Nielsen,Ruskai2002}.

The method employed in this work utilizes a quantum-classical analogy introduced in chapter \ref{sec:Analogia}.
Classical counterparts of quantum open systems are called stochastic systems \cite{hj2}.
Consequently, in terms of this analogy one can link general quantum channels with stochastic matrices
and unital quantum channels with bistochastic matrices.
In the standard approach bistochastic transformations represent the classical limit for quantum evolution \cite{duality2004,Landau,Grabert1979,Meara2013,Linial}.
However, in the present work we target in setting a non-asymptotic equivalence.

Apart from quantum considerations, the analysis of classical stochastic systems represent a worthwhile branch. Classical bistochastic matrices have been found very useful in practical applications, such as financial risk models \cite{Jarrow1997} or medical sciences \cite{Charitos2008}. Some preliminary results on stochastic roots of stochastic matrices of order three were reported in Ref.~\cite{He2003}. Nevertheless, ideas presented there need to be extended for our purposes.

Chapter \ref{sec:Analogia} explains on the qubit example, a peculiar link between the dynamics of quantum systems and stochastic systems.
Chapters \ref{sec:Parameterization} and \ref{sec:DynamicalSemigroups} refer to stochastic systems in the three-element space.
In chapter \ref{sec:Parameterization} we propose a parametrization for the set of $3\times3$ bistochastic matrices.
The following chapter \ref{sec:DynamicalSemigroups} contains a detailed analysis of the set of bistochastic matrices. In particular we target Markovian dynamics in stochastic systems.
Finally, in chapter \ref{sec:Parameterization4} we generalize the approach applied in chapter \ref{sec:Parameterization} to $4\times4$ matrices.

\section{Quantum-classical analogy}
\label{sec:Analogia}

The main goal of this paper is analysis of the time evolution of quantum open systems with three degrees of freedom. It is however convenient to introduce basic formalism and methodology on a simpler example of a \textit{qubit}, a quantum system with two degrees of freedom. This is the simplest quantum system capable of propagating information. Ideas presented for the qubit  are generalized in section \ref{sec:Parameterization} onto systems with three and four degrees of freedom.

\subsection{Description of a qubit}
The \textit{Qubit} can be described by the density operator in the two-dimensional Hilbert space \cite{Nielsen,duality2004}.
A choice of an orthonormal basis in the Hilbert space $\mathcal{H}$ provides a matrix representation for the density operator, called the density matrix.
For the \textit{qubit} the density matrix is given by
\begin{equation}
 \label{eq:postac_rho}
 \rho = \begin{pmatrix} p_1 & c \\ \bar c & p_2\end{pmatrix}
 \qquad
 \begin{array}{c}
  p_1,p_2\in\mathbb{R}_+ , ~ c \in \mathbb{C}, \\
  p_1 + p_2 = 1,\; p_1p_2 - |c|^2 \geq 0.
 \end{array}
\end{equation}

A change in state of the quantum system is determined by a map, which can be formally written as $\rho'=\mathcal{E}[\rho]$, and is called the \textit{quantum channel}  \cite{TheTheoryOfOpenQuantumSystems,Nielsen,duality2004,Landau}.
Here we consider a class of \textit{unital quantum channels} only, which in the case of \textit{qubits} is denoted as the class of \textit{Pauli channels} \cite{Ruskai2002}. 
Following Kraus \etal\ \cite{Kraus,Rivas2012}, the matrix representation for the \textit{Pauli channel} can be given by
\begin{equation}
 \label{def_operacji_Pauliego}
%  \mathbb{C}^{2\times2} \ni
 \rho
 \mapsto \mathcal{E}(\vec a)[\rho] = a_0 \rho + \sum_{\gamma\in\{x,y,z\}} a_\gamma \sigma_\gamma \rho \sigma_\gamma
%  \in \mathbb{C}^{2\times2},
 \qquad a_0 \equiv 1 - \sum_{\gamma\in\{x,y,z\}} a_\gamma ,
\end{equation}
where $\vec a = (a_x, a_y, a_z) \in\mathbb{R}_+^3$ are coefficients parametrizing the map (channel), and $\vec \sigma = (\sigma_x, \sigma_y, \sigma_z)$ is the vector of \textit{Pauli matrices}.

\begin{figure}[htbp]
 \centering
 \includegraphics[]{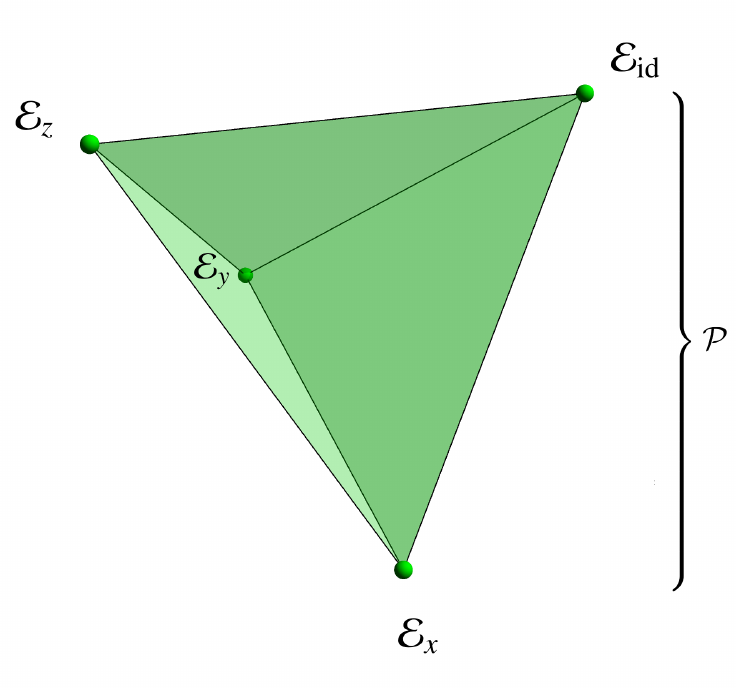}
 \caption{Geometrical representation of the set of all Pauli channels.  Vertices of the tetrahedron denoted as $\mathcal{E}_{\text{id}} , \mathcal{E}_{x}, \mathcal{E}_{y}, \mathcal{E}_{z}$ correspond to Pauli channels defined by eq. \ref{def_operacji_Pauliego}.}
 \label{fig:foliacja:bezfolii}
\end{figure}

The set of all \textit{Pauli channels}, denoted as $\mathcal{P}$, constitutes a tetrahedron which can be symbolically expressed as 
$\operatorname{span}( \mathcal{E}_{\text{id}} , \mathcal{E}_{x}, \mathcal{E}_{y}, \mathcal{E}_{z})$ \cite{GeomOfQState,Ruskai2002},
where: $\mathcal{E}_{\text{id}} [\rho] = \rho$ and $\mathcal{E}_\gamma[\rho] = \sigma_\gamma \rho \sigma_\gamma$ ($\gamma\in\{x,y,z\}$).
Figure \ref{fig:foliacja:bezfolii} displays an example of such tetrahedron. Motivation for a graphical representation of Pauli channels is supported by a classical-quantum analogy introduced in section \ref{sec:DynamicalSemigroups}.

\subsection{Classical part of description of a qubit}
\label{classical-quantum}
The state of the qubit can be represented by the density matrix or equivalently by a vector of \textit{classical probabilities} ($p_1$ and $p_2$)
completed with \textit{quantum coherences}~($c$). In the classical picture, quantum coherences are neglected, and the $\mathbf{p}$ vector fully characterizes the state of the qubit. This correspondence is marked with a wiggled arrow in eq. \ref{eq:QuantumClassicalAnalogy_state}.
\begin{equation}
 \label{eq:QuantumClassicalAnalogy_state}
 \rho 
%  = \begin{pmatrix} p_1 & c \\ \bar c & p_2\end{pmatrix}
 \rightsquigarrow
 \mathbf{p}
 =
 \begin{pmatrix} p_1 \\ p_2\end{pmatrix},
\end{equation}

Thus, for every Pauli channel, there exists an associated classical evolution $ \mathcal{E}[\rho] \rightsquigarrow \mathbf{B}\mathbf{p}$, which is determined by a stochastic matrix $\mathbf{B}$. Such an assignment is legitimized by the fact, that \textit{classical probabilities} characterizing the final state (after the evolution)
only depend on \textit{classical probabilities}, which refer to the initial state, i.e. classical probabilities of the final state are independent of the \textit{quantum coherences}.

\begin{figure}[htbp]
 \centering
 \includegraphics[]{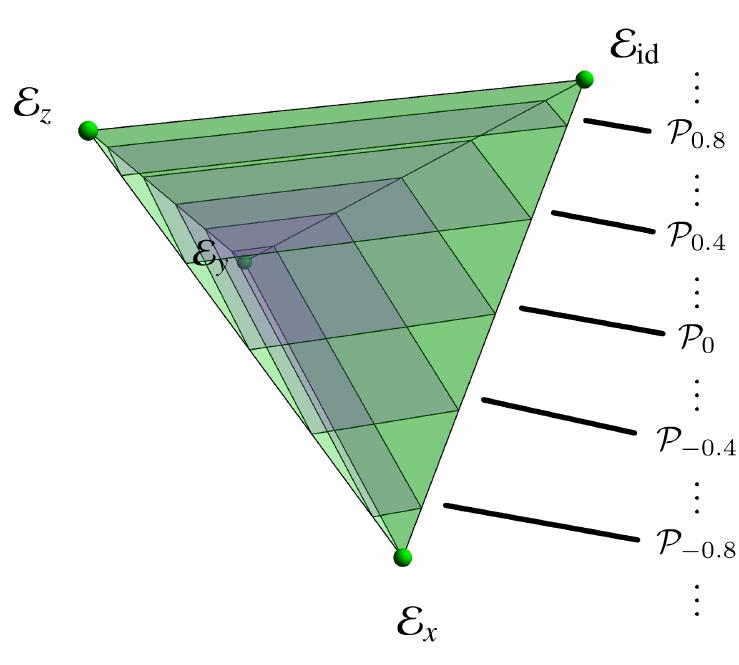}
 \caption{Foliation of the set of Pauli channels. Respective layers, denoted as $\mathcal{P}_\lambda$,  contain all quantum evolutions associated with single classical evolution.}
 \label{fig:foliacja:zfolia}
\end{figure}

A general form of~$\mathbf{B}$ is given by 
\begin{equation}
 \label{eq:identificationEB}
 \mathcal{E}[.]
 \rightsquigarrow
 \mathbf{B}
 =
 \begin{pmatrix}
  a_0 + a_z & a_x + a_y \\
  a_x + a_y & a_0 + a_z \\
 \end{pmatrix}
 .
\end{equation}
where $a_x, a_y, a_z $ are coefficients defining the Pauli channel in eq. \ref{def_operacji_Pauliego}. For this reason, classical bistochastic matrices are directly linked to the evolution of the quantum system. A chosen classical evolution can be realized by many quantum channels. This gives rise to an equivalence relation in the set of Pauli channels. Two Pauli channels are in a relation if and only if they generate identical classical evolution.
This equivalence relation imposes foliation of the set of Pauli channels.
Respective layers, denoted in Figure  \ref{fig:foliacja:zfolia} with $\mathcal{P}_\lambda$, contain all Pauli channels associated with a single bistochastic matrix. Eq. \ref{eq:foliation} presents the situation where $\lambda$ parametrizes a $2\times 2$ bistochastic matrix associated with the classical evolution: 

\begin{equation}
\label{eq:foliation}
 \mathcal{P}_\lambda
 \leftrightsquigarrow
 \mathbf{B}(\lambda)
 =
 \frac{1}{2} \begin{pmatrix} 1+\lambda & 1-\lambda \\ 1-\lambda & 1+\lambda \end{pmatrix}
 .
\end{equation}

\subsection{Dynamics}

In general, the classical picture of time evolution of the system is far more intuitive and more accessible, than the full quantum description. At the same time, quantum dynamics of a number of systems can be analysed and interpreted by means of their classical properties. As shown in section \ref{classical-quantum}, the qubit is an example of such system. Thus, in the present section we will focus on the classical part of the dynamics of the qubit. From this perspective, the qubit is a two dimensional classical stochastic system, which has only one dynamical semigroup associated with it,
\begin{equation}
 \label{eq:BiSt22Range_izo}
 \left\{
   \frac{1}{2} \begin{pmatrix} 1 + e^{-v_z t} & 1 - e^{-v_z t}\\ 1 - e^{-v_z t} & 1 + e^{-v_z t}\end{pmatrix}
 \middle|
  t\in \mathbb{R}_+
 \right\},
\end{equation}
where $v_z > 0$ is a parameter, which identifies the parametrization of the unique dynamical semigroup. This parametrization can be viewed as an isomorphism between the semigroup and $(\mathbb{R}_+,+)$. A necessary condition for a group of quantum evolutions to be well defined, is the existence of a group of classical time evolutions associated with the quantum group. In the present case, this means that the quantum group (parametrized by $\vec a(t)$) has to satisfy the following condition:
\begin{equation}
 \label{eq:comesDown_matrix_condition}
 \begin{pmatrix}
  a_0(t) + a_z(t) & a_x(t) + a_y(t) \\
  a_x(t) + a_y(t) & a_0(t) + a_z(t)
 \end{pmatrix}
 =
 \frac{1}{2}
 \begin{pmatrix}
  1 + e^{-v_z t} & 1 - e^{-v_z t} \\
  1 - e^{-v_z t} & 1 + e^{-v_z t}
 \end{pmatrix} ;
\end{equation}
which, can be rewritten in a compact scalar form
\begin{equation}
 \label{eq:comesDown_scalar_condition} 
 a_0(t) + a_z(t) = \frac{1}{2} \big( 1 + e^{-v_z t} \big) .
\end{equation}
From the geometric point of view, eq. \ref{eq:comesDown_scalar_condition} provides information about the dynamics in the resolution of a single layer. Dynamics inside the layer is completely undetermined. 

The set of Pauli channels is a specific case, where it is straightforward to provide a general formula for all Markov groups of time evolutions.
\begin{equation}
 \Big\{ \mathcal{E}\big( \vec a(\vec v; t))[.] \Big| t\in \mathbb{R}_+ \Big\}
\end{equation}
where
\begin{equation}
 \label{eq:PauliAfine_izo}
  \begin{pmatrix}  a_x(\vec v; t) \\ a_y(\vec v; t) \\ a_z(\vec v; t) \end{pmatrix}
  =
  \frac{1}{4}
  \begin{pmatrix} 1 \\ 1 \\ 1 \end{pmatrix}
  +
  \frac{1}{4}
  \begin{pmatrix} +1 & -1 & -1 \\ -1 & +1 & -1 \\ -1 & -1 & +1 \end{pmatrix}
  \begin{pmatrix} e^{-v_x t} \\ e^{-v_y t} \\ e^{-v_z t} \end{pmatrix}  .
\end{equation}
Direction of the $\vec v$ vector identifies the semigroup, and the length of $\vec v$ determines the parametrization of the semigroup - an isomorphism between the semigroup and $(\mathbb{R}_+,+)$. In other words, the set of all dynamical semigroups is two dimensional, and can be parametrized by the unit vector $\vec v / |\vec v|$. Comparison of eq. \ref{eq:PauliAfine_izo} and eq. \ref{eq:BiSt22Range_izo}, with the help of the formula in eq. \ref{eq:identificationEB}, suggests that $v_z$ introduced in eq. \ref{eq:BiSt22Range_izo} and $v_z$ introduced in eq. \ref{eq:PauliAfine_izo} refer to the same quantity. This means that the $v_z$ component of the vector $\vec v$ is directly linked to the classical evolution. As a consequence, the dynamics of quantum channels can be investigated by means of the classical-quantum equivalence presented above.
In the first step, the classical-quantum equivalence is established, as it was done in eq.~\ref{eq:identificationEB}. At this stage, a choice of the quantization axis is required (here, the $z$-direction was chosen). In the next step, a detailed analysis of classical Markov semigroups is performed (see eq.~\ref{eq:BiSt22Range_izo}).
Equations mentioned in steps one and two allow to derive a condition for a
quantum evolution to be consistent with an associated classical evolution (see eq.~\ref{eq:comesDown_scalar_condition}).

The consistency condition is insufficient for complete reproduction of the quantum dynamics in $\mathcal{P}$.
Nonetheless, it provides a constraint on Markov semigroups existing in $\mathcal{P}$. 
The condition is associated with the chosen basis in the Hilbert space. Fortunately, one is free to chose different bases.
For example, in the case of the Pauli channels, the choice of three different quantization axis $x$, $y$ and $z$ results in three independent conditions,
which if solved, deliver a general formula for any semigroup in $\mathcal{P}$.

\section{Mathematical objects and notation}

\subsection{Motivation}
The quantum-classical equivalence introduced in the previous section, was shown to provide a simple way for describing the dynamics of the quantum system, with the use of 
classical dynamical groups. The form of the classical dynamical group for the qubit can be straightforwardly postulated, as done in eq. \ref{eq:BiSt22Range_izo}, without any derivation, due to simplicity of the two-dimensional system. 
The knowledge of dynamical groups in this generic case allowed to conclude, that unital quantum evolutions correspond to subgroups of the group of bistochastic matrices.
In systems with more than two degrees of freedom, the explicit form of the classical dynamical groups requires a more rigorous approach, which is developed in this section.

\subsection{Introduction of mathematical objects}

For the sake of self-consistency, key mathematical objects are briefly introduced below.

The set of all bistochastic matrices of size $N$ is denoted with $\mathcal{B}_N$. Bistochastic matrices from $\mathcal{B}_N$ describe transformations of \textit{classical stochastic systems} with $N$ degrees of freedom, for which the maximum entropy state is the steady state. This fact puts constraints on bistochastic matrices, written as

\begin{den}[Sets $\mathcal{B}_N$ for $N \in \mathbb{N}$]
 \begin{align}\label{bistochastic}
  \mathcal{B}_N
  &=
  \bigg\{ \mathbf{B} \in \mathbb{R}^{N\times N} \bigg| \sum_i B_{ij}=1, ~ \sum_j B_{ij}=1, ~B_{ij}\geq0 \bigg\} .
 \end{align}
\end{den}

The geometry of the $\mathcal{B}_N$ set is analysed in the Birkhoff--von Neumann theory \cite{Birkhoff}, where $\mathcal{B}_N$  is represented as a polyhedron, which constitutes convex hull of the set of $N!$ permutation matrices.  This polyhedron is also called the $N$-dimensional Birkhoff polytope \cite{Landau,CourInConv}.
$\mathcal{B}_N$ forms a group, and in this work, we will focus on dynamical subgroups of $\mathcal{B}_N$, i.e. group structures in the interior of the Birkhoff polytope.

Definition in eq. \ref{bistochastic} contains two conditions: summation to unity of elements in every column and every row,  and non-negativity of individual matrix elements. Here, it is useful to introduce an auxiliary definition, which contains only the former condition:

\begin{den}[Sets $\mathcal{W}_N$ for $N \in \mathbb{N}$]
 \begin{align}\label{bistochastic2}
  \mathcal{W}_N
  &=
  \bigg\{ \mathbf{W} \in \mathbb{R}^{N\times N} \bigg| \sum_i W_{ij}=1, ~ \sum_j W_{ij}=1 \bigg\} ,
 \end{align}
\end{den}

Let us denote with $\mathcal{B}_{N\text{sym}}$ and $\mathcal{W}_{N\text{sym}}$, respective sets of symmetric matrices in $\mathcal{B}_N$ and $\mathcal{W}_N$.
Below we define subsets of $\mathcal{B}_N$, which will be used further on.

\begin{den}(set $\mathcal{B}_{3\text{sym}}^\text{Markov}$,
            set $\mathcal{B}_{3\text{sym}}^{\text{MarkovLimit}}$
            and set $\mathcal{B}_{3\text{sym}}^{\infty \text{root}}$)
We denote:
\begin{itemize}
 \item the set of matrices describing classical Markov evolution:
 \begin{equation}
  \mathcal{B}_{3\text{sym}}^\text{Markov}
  =\\
  \Big\{
  e^{\mathbf{L}} \in \mathcal{B}_{3\text{sym}}
  \Big|
  \mathbf{L} \in \mathbb{R}^{3\times3} :
  \big\{ e^{t\mathbf{L}} \big| t>0 \big\} \subset \mathcal{B}_{3\text{sym}}
  \Big\}  ;
 \end{equation}

 \item the set of matrices describing Markov evolution extended with feasible asymptotic evolutions (infinite time evolutions):
 \begin{equation}
  \mathcal{B}_{3\text{sym}}^{\text{MarkovLimit}}
  =
  \mathcal{B}_{3\text{sym}}^\text{Markov}
  \bigcup
  \Big\{
  \lim_{t\rightarrow\infty} e^{t\mathbf{L}} \in \mathcal{B}_{3\text{sym}}
  \Big|
  \mathbf{L} \in \mathbb{R}^{3\times3} :
  \big\{ e^{t\mathbf{L}} \big| t>0 \big\} \subset \mathcal{B}_{3\text{sym}}
  \Big\}  ;
 \end{equation}

 \item the set of divisible bistochastic matrices:
 \begin{equation}
  \mathcal{B}_{3\text{sym}}^{\infty \text{root}}
  =
  \left\{
  \mathbf{B} \in \mathcal{B}_{3\text{sym}}
  \middle|
  \begin{array}{c}
   \forall n\in\mathbb{N}   ~    \exists \mathbf{B}_n \in \mathcal{B}_{3\text{sym}}  :  \\
   \mathbf{B} = (\mathbf{B}_n)^n
  \end{array}
  \right\} .
 \end{equation}
 \end{itemize}
\end{den}

\section{Parametrization of the Birkhoff polytope of order 3}
\label{sec:Parameterization}

\subsection{Definition of the parametrization}

A $3\times3$ matrix contains nine independent elements. However, the condition in eq.~\ref{bistochastic2}, narrows the number of parameters for a $3\times3$ matrix to four independent elements.
For this reason, a more convenient representation of the bistochastic matrix can be obtained by a suitable parametrization. The choice of such parametrization should lead to  an isomorphism with another group, which has a simpler structure. Here we postulate, the geometry of the set of bistochastic matrices is related to the geometry of complex numbers. Accordingly, a complex parametrization of the set of bistochastic matrices can be proposed:

\begin{obs}[Parametrization of the $\mathcal{W}_3$ set with complex numbers $u,w$]
 \label{obs:parametryzacja}
 Every matrix contained in $\mathcal{W}_3$ can be expressed as
 \begin{equation}
  \label{FormulaMaster}
  \mathbf{W}(u,w)
  = \mathbf{B}_\star
  + \frac{2}{3} \operatorname{Re} \left[ u \mathbf{M}_1 \right]
  + \frac{2}{3} \operatorname{Re} \left[ w \mathbf{M}_2 \right]
 \end{equation}
 for any $u \in \mathbb{C}, w \in \mathbb{C}$, with the respective matrices defined as below
 \begin{equation*}
   \mathbf{B}_\star   = \frac{1}{3} \begin{pmatrix} 1 & 1 & 1 \\ 1 & 1 & 1 \\ 1 & 1 & 1 \end{pmatrix} , \qquad
   \mathbf{M}_1 = \begin{pmatrix} 1 & \Omega & \Omega^2 \\ \Omega^2 & 1 & \Omega \\ \Omega & \Omega^2 & 1 \end{pmatrix} , \qquad
   \mathbf{M}_2 = \begin{pmatrix} \Omega^2 & 1 & \Omega \\ 1 & \Omega & \Omega^2 \\ \Omega & \Omega^2 & 1 \end{pmatrix}
 \end{equation*}
 where
 $\Omega = \exp\big(i\tfrac{2}{3}\pi\big)$.
\end{obs}

Equation \ref{FormulaMaster} establishes 1:1 correspondence between the set $\mathcal{W}_3$ and pairs $(u,w) \in \mathbb{C}^2$.
Table \ref{tab:special_points} lists appropriate $(u,w)$ values for six permutation matrices -- vertices of the Birkhoff polytope of order 3.
\begin{table}[h]
\begin{displaymath}
\begin{array}{|cc|cc|}
 \hline
 \text{matrix} & \text{coordinate} &
 \text{matrix} & \text{coordinate}  \\
 \hline
 \hline
 \begin{array}{c} \mathbf{P}_e =\\= \mathbf{W}(1,0) \end{array}
 &
 \left\{
 \begin{array}{l} u=1 \\ w=0 \end{array}
 \right.
 &
 \begin{array}{c} \mathbf{P}_{(12)} =\\= \mathbf{W}(0,1) \end{array}
 &
 \left\{
 \begin{array}{l} u=0 \\ w=1 \end{array}
 \right.
 \\ \hline
 \begin{array}{c} \mathbf{P}_{(123)} =\\= \mathbf{W}(\Omega^2,0) \end{array}
 &
 \left\{
 \begin{array}{l} u=\Omega^2 \\ w=0 \end{array}
 \right.
 &
 \begin{array}{c} \mathbf{P}_{(13)} =\\= \mathbf{W}(0,\Omega^2) \end{array}
 &
 \left\{
 \begin{array}{l} u=0 \\ w=\Omega^2 \end{array}
 \right.
 \\ \hline
 \begin{array}{c} \mathbf{P}_{(132)} =\\= \mathbf{W}(\Omega,0) \end{array}
 &
 \left\{
 \begin{array}{l} u=\Omega \\ w=0 \end{array}
 \right.
 &
 \begin{array}{c} \mathbf{P}_{(23)} =\\= \mathbf{W}(0,\Omega) \end{array}
 &
 \left\{
 \begin{array}{l} u=0 \\ w=\Omega \end{array}
 \right.
 \\ \hline
 \end{array}
\end{displaymath}
\caption{Collection of permutation matrices and their complex coordinates. We denote $\Omega = \exp\big(i\tfrac{2}{3}\pi\big)$.}
\label{tab:special_points}
\end{table}

\subsection{Properties of the parametrization}
The main advantage of the chosen parametrization is a significant simplification of analysis of the Birkhoff polytope.
Indeed, nine real matrix elements are transformed into four independent real parameters arranged in a pair of complex numbers. 
These pairs can be further arrayed into $2\times2$ matrices, which form a representation of the $(\mathcal{W}_3,\cdot)$ group.
\begin{property}[Two-dimensional representation] An isomorphism defined as
 \begin{equation}
  \mathbf{W}(u,w)
  \sim
  \begin{pmatrix}
   u & \bar w \\
   w & \bar u \\
  \end{pmatrix} .
 \end{equation}
 \label{prop:2Dw3}
 constitutes a two-dimensional representation of the $(\mathcal{W}_3,\cdot)$ group.
\end{property}
Below we list characteristics of selected important subgroups of $\mathcal{W}_{3\text{sym}}$. 

\begin{property}[Classification of the $\mathcal{W}_{3\text{sym}}$ subgroup in the framework of the complex parametrization $(u,w)$]
 \label{prop:Charakt_Mac_sym}
 $\mathbf{W}(u,w) \in \mathcal{W}_{3\text{sym}} \Leftrightarrow u \in \mathbb{R}$.
\end{property}

Let us define half-planes in $\mathcal{W}_{3\text{sym}}$, which are closed under multiplication of matrices. The analysis of the dynamical semigroups in $\mathcal{B}_{3\text{sym}}$ is largely simplified in terms of these half-planes.
\begin{property}[Half-planes closed under multiplication]
 \label{prop:half-planes}
 There exist two-dimensional subsets of $\mathcal{W}_{3\text{sym}}$ closed under multiplication of matrices.
 These subsets define half-planes $\mathcal{W}_\phi$, $\phi \in [0,2\pi)$:
 \begin{equation}
  \mathcal{W}_\phi = \big\{ \mathbf{W} (u,w) \big| u\in\mathbb{R}, \operatorname{arg} (w) = \phi \big\}.
 \end{equation}
 Of course $\bigcup \mathcal{W}_\phi = \mathcal{W}_{3\text{sym}}$.
\end{property}
As a consequence, the half plane $\mathcal{W}_\phi$ can be described in the two-dimensional space of parameters $\big(u, |w| \big) \in \mathbb{R} \times \mathbb{R}_+$.
Within the introduced half-planes $\mathcal{W}_\phi$ two regions are of special importance: a region of positive definite matrices and a region of bistochastic matrices. Both regions are characterized by the following relations:

\begin{property} Characterization of the set of positive definite matrices 
 \begin{equation}
  \mathcal{W}_\phi \cap  \big\{ \mathbf{M} \in \mathbb{R}^{3\times3} \big| \mathbf{M} >0 \big\} =
  \big\{ \mathbf{W} (u, |w| e^{i\phi}) \big| u > |w| \big\} .
 \end{equation}

\end{property}

\begin{property} Characterization of the set of bistochastic matrices 
 \begin{equation}
 \label{eq:MarkovInW}
  \mathcal{W}_\phi \cap \mathcal{B}_{N\text{sym}}
  =
  \operatorname{span}
   \Big\{
     \mathbf{W} \big(-1/2,0\big) ,
     \mathbf{W} \big(1,0\big),
     \mathbf{W} \big(0,f(\phi)e^{i\phi} \big)
   \Big\}
 \end{equation}

 where $f(\phi)$ is an auxiliary function defined as
 \begin{equation}
  \label{eq:fOdPhi}
   f(\phi) =
   \left\{
    \begin{array}{ll}
               +\frac{1}{2}\sec(\phi-\pi/3) & \text{for: } \phi \in [0, 2\pi/3], \\
               -\frac{1}{2}\sec(\phi)   & \text{for: } \phi \in [2\pi/3, 4\pi/3],  \\
               +\frac{1}{2}\sec(\phi+\pi/3) & \text{for: } \phi \in [4\pi/3, 2\pi].
    \end{array}
   \right.
 \end{equation}
\end{property}

\subsection{Graphical representation}

\begin{figure}[t!]
  \centering
  \includegraphics[]{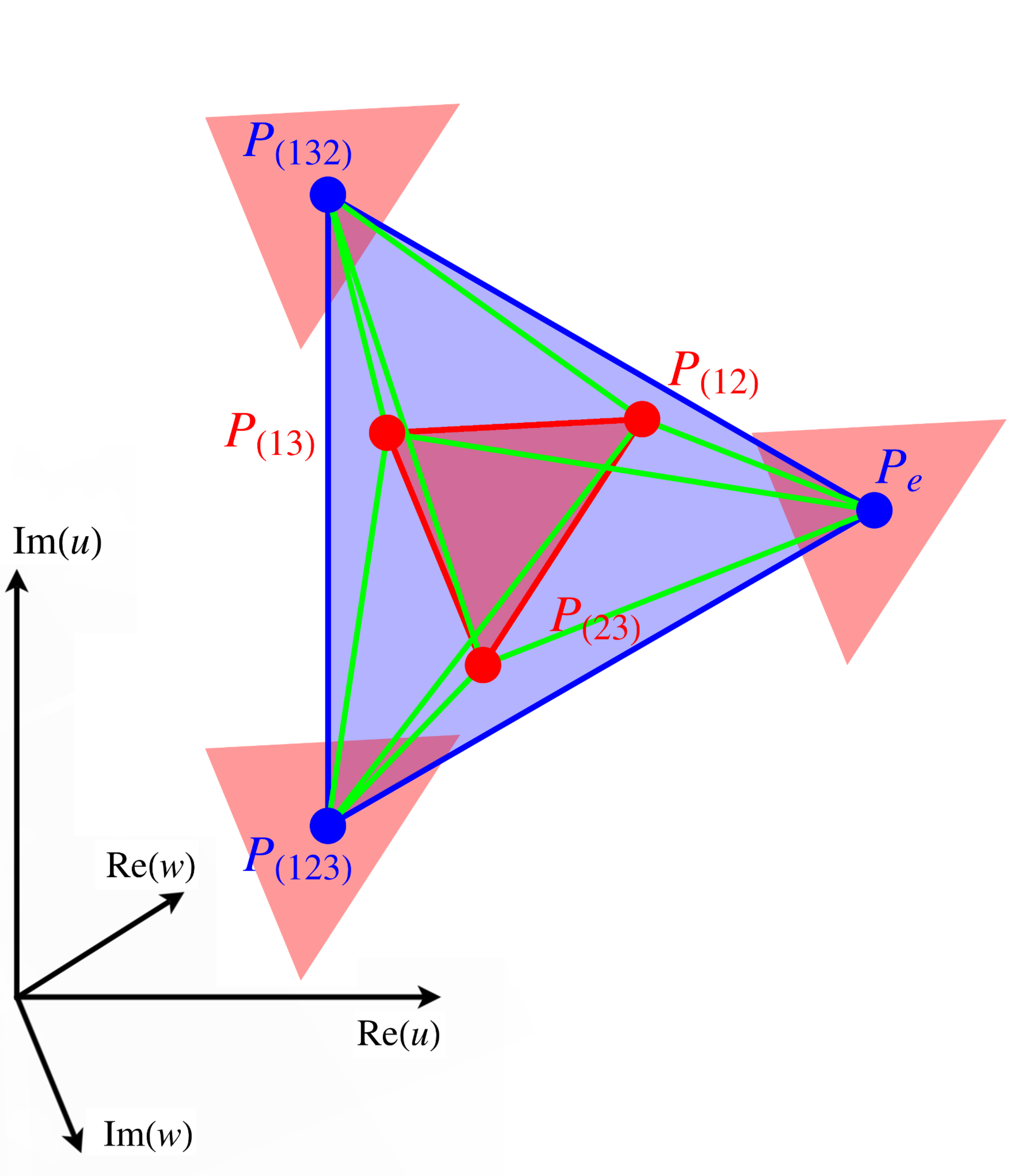} 
  \caption{Representation of the Birkhoff polytope in terms of the  parametrization from observation \ref{obs:parametryzacja}.
	   The blue triangle $(P_e,P_{(123)},P_{(132)})$ is spanned by points with constant $w$ coordinate ($w=0$),
           while red triangles represent points with constant $u$ coordinate
           ($u=0$, $u=1$, $u=\Omega$, $u=\Omega^2$ respectively).
           Vertices of the triangles correspond to permutation matrices.
           Edges of the polytope are marked with three blue, three red and six green lines.
           (Note that the origin of the coordinate system is located in the center of the blue triangle).}
  \label{fig:tesseract_one}
\end{figure}

The complex parametrization given in observation \ref{obs:parametryzacja} enables a simple graphical representation of the Birkhoff polytope. Indeed, the parametrization conserves the structure of the affine combination of points in the Birkhoff polytope, thus any polytope contained in $\mathcal{W}_3$ corresponds to a polytope in $\mathbb{C}^2$. Vertices of the former and the latter polytope overlap. For this reason, the Birkhoff polytope is represented by a polytope in $\mathbb{C}^2$, which is spanned by points collected in Table \ref{tab:special_points}. This situation is displayed in Figure \ref{fig:tesseract_one}.

\begin{figure}[t!]
 \centering
  \includegraphics{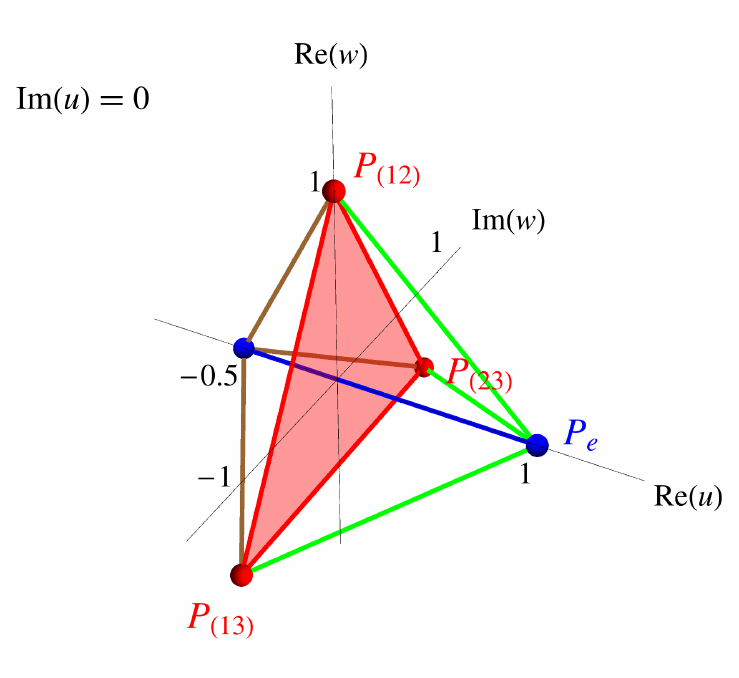} 
  \caption{Graphical representation of $\mathcal{B}_{3\text{sym}}$.
  The color code used in this figure is consistent with the color code used in fig \ref{fig:tesseract_one}.
  Three green segments, three red segments and three brown segments are the edges of the trigonal bipiramid representing $\mathcal{B}_{3\text{sym}}$.
  All of them but the brown segments are also edges of the polygon representing $\mathcal{B}_{3}$.
  }
  \label{fig:BirkhoffSym}
\end{figure}

As shown in property $\ref{prop:Charakt_Mac_sym}$, symmetric bistochastic matrices are represented by points $(u,w) \in \mathbb{R}\times\mathbb{C}$. Then, 
$\mathcal{B}_{3\text{sym}}$ can be visualized as the intersection of the polytope shown in Figure $\ref{fig:tesseract_one}$ with $\mathbb{R}\times\mathbb{C}$.
In other words, the representation of $\mathcal{B}_{3\text{sym}}$ is directly obtained from the polytope in Figure $\ref{fig:tesseract_one}$,
by neglecting the $\text{Im}(u)$ dimension. In this way, $\mathcal{B}_{3\text{sym}}$ is represented by the trigonal bipiramid, whereas
vertices of $\mathcal{B}_{3\text{sym}}$ correspond to the permutation matrices $\mathbf{P}_{e}, \mathbf{P}_{(13)},\mathbf{P}_{(12)}, \mathbf{P}_{(23)}$ and  $\tfrac{1}{2}(\mathbf{P}_{(123)}+\mathbf{P}_{(132)})$ (see Figure \ref{fig:BirkhoffSym}).

\begin{figure}[t!]
 \centering
 \includegraphics{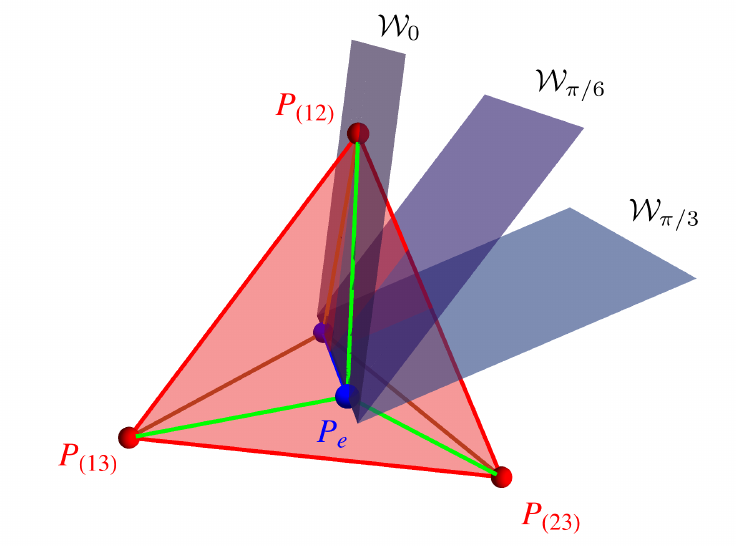} 
 \caption{Representation of example half-planes $\mathcal{W}_\phi$ (corresponding to $\phi$ equal to $0$, $\pi/6$ and $\pi/3$)
 together with the trigonal bipiramid representing $\mathcal{B}_{3\text{sym}}$.}
 \label{fig:BsymPhiPlaneS}
\end{figure}

Example half-planes ($\mathcal{W}_{0}$, $\mathcal{W}_{\pi/6}$, $\mathcal{W}_{\pi/3}$),  defined by property~\ref{prop:half-planes}, are displayed in  Figure \ref{fig:BsymPhiPlaneS}. These three half-planes will be used in the next section in discussion of the dynamical semigroups.

\section{Dynamical semigroups in the set of bistochastic matrices of order 3}
\label{sec:DynamicalSemigroups}

\subsection{Symmetric matrices case}

Every dynamical semigroup in $\mathcal{W}_{3\text{sym}}$ belongs to a half-plane $\mathcal{W}_\phi$. It is therefore convenient to formulate results in terms of dynamical semigroups in an arbitrary half-plane $\mathcal{W}_\phi$, as done in eq. \ref{eq:MarkovSemigroupsInWphi}.

\begin{lem}[Markov semigroups in $\mathcal{W}_\phi$]
$\mathcal{W}_\phi$ includes the following semigroups
\begin{equation}
  \label{eq:MarkovSemigroupsInWphi}
  \Big\{
   \mathbf{W}\big( u(\theta;t) , \big|w(\theta;t)\big| e^{i\phi} \big)
  \Big|
   t \in [0,\infty)
  \Big\}
  \subset \mathcal{W}_\phi
\end{equation}
parametrized as follows
\begin{equation}
 \left\{
 \begin{array}{rcl}
  u(\theta;t)            &=& \operatorname{exp}(-t\cos\theta) \cosh(t\sin\theta)  \\
  \big|w(\theta;t)\big|  &=& \operatorname{exp}(-t\cos\theta) \sinh(t\sin\theta)  \\
 \end{array}
 \right.
\end{equation}
where $\theta\in[0,\pi]$ is a parameter characterizing the semigroup.
\end{lem}

\begin{figure}[htbp]
 \centering
 \includegraphics{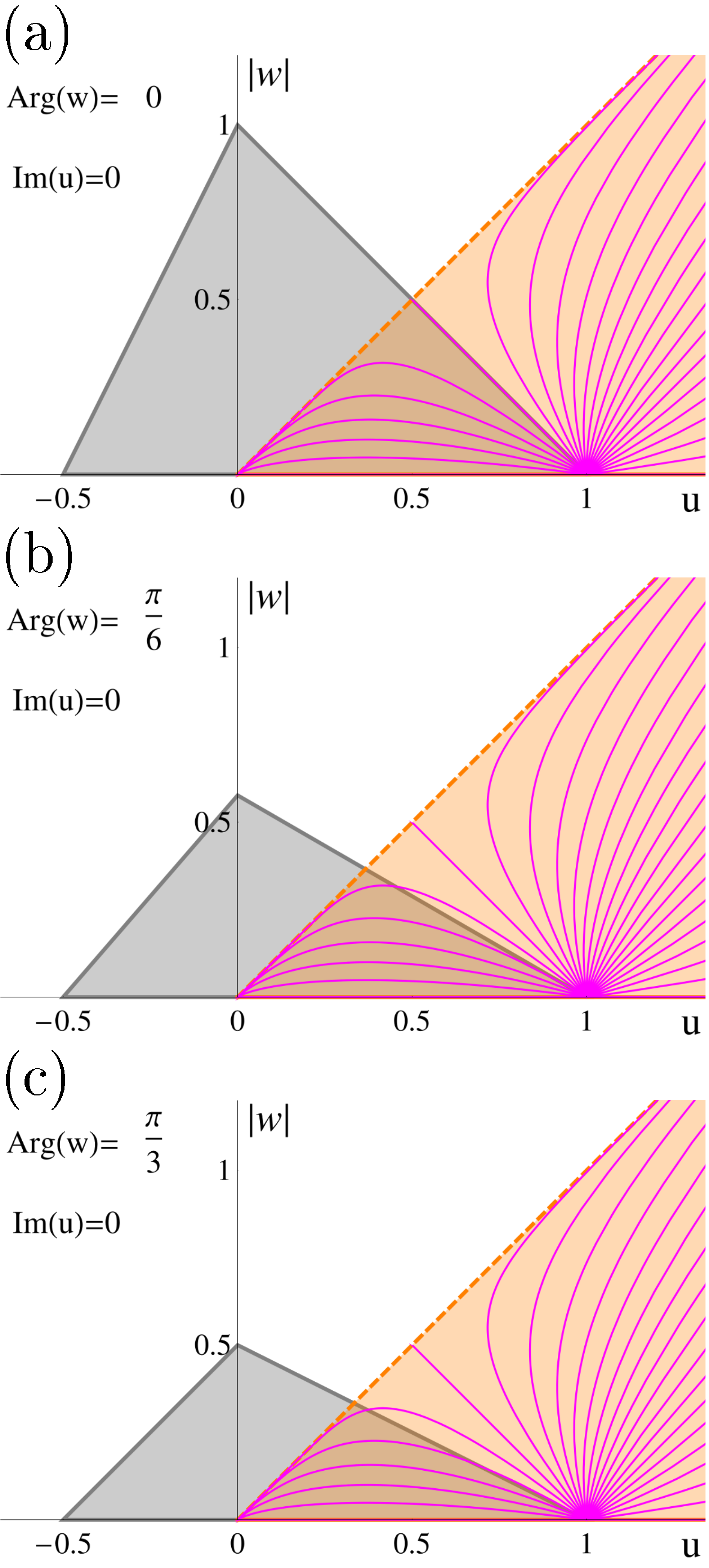}
 \caption{Analysis of half-planes $\mathcal{W}_\phi$ for (a) $\phi=0$, (b) $\phi=\pi/6$ and (c) $\phi=\pi/3$.
          Grey shadowed region represents the subset of bistochastic matrices,
          and the orange region (with dashed edge) encloses the subset of all positive definite matrices.
          Purple lines in the orange region give examples of one-parameter semigroups in $\mathcal{W}_{3\text{sym}}$.}
 \label{fig:Wphi}
\end{figure}

A semigroup characterized by the $\theta$ parameter is tangent (for its identity representative)
to a segment connecting identity (unit matrix) with $\mathbf{W}( 1-\cos\theta , \sin\theta e^{i\phi} )$. Examples of such segments are depicted in Figure \ref{fig:Wphi}.

Note that both $u(\theta;t)$ and $\big|w(\theta;t)\big|$ are independent of the complex argument $\phi$.
Thus, the considered semigroups conserve full rotational symmetry along the axis defined by affine combinations of the $\mathbf{B}_\star$ and $\mathbf{P}_e$ matrices.
This symmetry is reduced to the three fold axis for the $\mathcal{B}_3$ set.

\subsection{Characterization of $\mathcal{B}_{3\text{sym}}^{\text{MarkovLimit}}$ and $\mathcal{B}_{3\text{sym}}^{\infty \text{root}}$}

The previous section delivered tools necessary to formulate a criterion for a matrix belonging to the semigroup of bistochastic symmetric Markov $3\times3$ matrices.

\begin{OurCorollary}(Criterion for the $\mathbf{W}(a,b e^{i\phi})$,
 $a\in\mathbb{R}$, $b\in\mathbb{R}_+$, $\phi\in[0,2\pi)$ matrix to be contained in $\mathcal{B}_{3\text{sym}}^{\text{MarkovLimit}}$).

\begin{enumerate}
  \item if $a<b$, then $\mathbf{W}(a,b e^{i\phi})$ does not belong to
  any of the semigroups described by the formula \textcolor[rgb]{0.00,0.00,0.59}{ \ref{eq:MarkovSemigroupsInWphi} }.
  In this case $\mathbf{W} \not\in \mathcal{B}_{3\text{sym}}^{\text{MarkovLimit}}$.
  \item if $a>b$, then the  $\mathbf{W}(a,b e^{i\phi})$ matrix belongs to the semigroup described by the formula \textcolor[rgb]{0.00,0.00,0.59}{ \ref{eq:MarkovSemigroupsInWphi}}.
   The $\theta$ parameter describes a semigroup containing matrices which satisfy one of the conditions above and is given by the following formula:
  \begin{equation}
   \tan \theta
   %= -2
   %  \frac
   %  { \operatorname{artgh} \big( \frac{b}{a} \big) }
   %  { \ln \big( a^2 - b^2 \big)  }
  =
  \frac { \ln(a-b) - \ln(a+b) }
        { \ln(a-b) + \ln(a+b) }
   .
  \end{equation}
  Limitation on $\tan \theta$:
  \begin{equation}
    0 \leq \tan \theta \leq f(\phi)
  \end{equation}
  implies $\mathbf{W} \in \mathcal{B}_{3\text{sym}}^{\text{MarkovLimit}}$.
  In the opposite case: $\mathbf{W} \not\in \mathcal{B}_{3\text{sym}}^{\text{MarkovLimit}}$.

  \item in the $a=b$ case, the $\mathbf{W}(a,b e^{i\phi})$ matrix belong to the semigroup described by the formula \textcolor[rgb]{0.00,0.00,0.59}{ \ref{eq:MarkovSemigroupsInWphi} } only if $(a,b)=(0,0)$ or $(a,b)=(\tfrac{1}{2},\tfrac{1}{2})$.
  Hence, $\mathbf{W} \in \mathcal{B}_{3\text{sym}}^{\text{MarkovLimit}}$
  in four exclusive cases:
  \begin{equation}
  \begin{aligned}
   &(a,b) = (0,0), &
   &(a,b,\phi) = \big( \tfrac{1}{2}, \tfrac{1}{2}, 0 \big) , \\
   &(a,b,\phi) = \big( \tfrac{1}{2}, \tfrac{1}{2}, \tfrac{2}{3}\pi \big),&
   &(a,b,\phi) = \big( \tfrac{1}{2}, \tfrac{1}{2}, \tfrac{4}{3}\pi \big).
  \end{aligned}
  \end{equation}
\end{enumerate}
\end{OurCorollary}

The above conditions provide a procedure for finding matrices, which describe classical Markov evolution of 3-state systems.
Note that the set of infinitely divisible matrices $\mathcal{B}_{3\text{sym}}^{\text{MarkovLimit}}$ is included in $\mathcal{B}_{3\text{sym}}^{\infty \text{root}}$, but inclusion in the opposite direction is not generally satisfied. The difference between  $\mathcal{B}_{3\text{sym}}^{\text{MarkovLimit}}$
and $\mathcal{B}_{3\text{sym}}^{\infty \text{root}}$ is quite subtle. Interiors of the two sets are identical. At the boundaries,
$\mathcal{B}_{3\text{sym}}^{\infty \text{root}}$ is a closed set, unlike $\mathcal{B}_{3\text{sym}}^{\text{MarkovLimit}}$. More precisely, the difference between these sets  occurs for three segments:

\begin{equation}
\label{difference}
  \mathcal{B}_{3\text{sym}}^{\infty \text{root}}
  \setminus
  \mathcal{B}_{3\text{sym}}^{\text{MarkovLimit}}
  =
 \bigcup_{\phi \in \{0,\frac{2\pi}{3},\frac{4\pi}{3}\}}
 \left\{
  \mathbf{W}(x,xe^{i\phi})
 \middle|
  x \in \bigg(0,\frac{1}{2} \bigg)
 \right\}
 .
\end{equation}

The relation in eq. \ref{difference} is rationalized by the observation that $\mathcal{B}_{3\text{sym}}$ contains one-parameter semigroups, in which the neutral element is a non-unit matrix, i.e.
 \begin{equation}
  \bigg\{
   \mathbf{W}\bigg( \frac{1}{2} e^{-t} , \frac{1}{2} e^{-t} e^{i\phi} \bigg)
  \bigg|
   t \in [0,\infty)
  \bigg\}
  \subset \mathcal{B}_{3\text{sym}} \\
 \end{equation}
for  $\phi = \big\{ 0,\frac{2\pi}{3},\frac{4\pi}{3} \big\}$,
with the neutral element given by $\mathbf{W}\big(\frac{1}{2},\frac{1}{2}e^{i\phi} \big)$.
Each of these subgroups is associated to a single segment in
$\mathcal{B}_{3\text{sym}}^{\infty \text{root}} \setminus \mathcal{B}_{3\text{sym}}^{\text{MarkovLimit}}$.

At the end of this section, let us introduce a relation between $\mathcal{B}_{3\text{sym}}^\text{Markov}$ and the set of positive $3\times3$ matrices.
Figure \ref{fig:BsymPhiPlaneS} shows that: 
$\mathcal{B}_{3\text{sym}}^\text{Markov}
 \subsetneq
 \big\{ \mathbf{B} \big| \mathbf{B} > 0 \big\}$
and
$\mathcal{B}_{3\text{sym}}^{\text{MarkovLimit}}
 \subsetneq
 \big\{ \mathbf{B} \big| \mathbf{B} \geq 0 \big\}$.

However, it is possible to find symmetric bistochastic matrices, which are simultaneously
(a) contained in the Markov semigroup (included in the Birkhoff polytope) and
(b) not positive definite.
Below is an example of such matrix:
 \begin{equation}
  \exp
  \left[
   \frac{2}{\sqrt{3}}
   \left(
   \begin{smallmatrix}
    -1 &  0 & +1 \\
    +1 & -1 &  0 \\
     0 & +1 & -1
   \end{smallmatrix}
   \right)
   t
  \right]
  =
  \mathbf{W} \Big(e^{-\sqrt{3}t}e^{it}, 0\Big)
  \in
  \mathcal{B}_{3}
 \end{equation}
 together with its representative for $t:=\pi$
 \begin{equation}
  \frac{1}{3}
  \left(
  \begin{smallmatrix}
   1 - 2e^{-\sqrt{3} \pi } & 1 + e^{-\sqrt{3} \pi }  & 1 + e^{-\sqrt{3} \pi }  \\
   1 + e^{-\sqrt{3} \pi }  & 1 - 2e^{-\sqrt{3} \pi } & 1 + e^{-\sqrt{3} \pi }  \\
   1 + e^{-\sqrt{3} \pi }  & 1 + e^{-\sqrt{3} \pi }  & 1 - 2e^{-\sqrt{3} \pi } \\
  \end{smallmatrix}
  \right)
  =
  \mathbf{W} \big(-e^{-\sqrt{3}\pi}, 0\big)
  \in
  \mathcal{B}_{3\text{sym}} .
 \end{equation}

\subsection{General case}

The $\mathcal{B}_{3}$ set is characterized by four independent parameters, which means that the 4-dimensional geometry of this set can only be partially visualized (see Figure \ref{fig:tesseract_one}). A graphical representation for one-parameter Markov semigroups occurs even more challenging due an additional parameter. Nonetheless, in the introduced framework of the complex parametrization this issue becomes largely simplified. A support for this statement is given by the following lemma:

\begin{lem}[Markov semigroups in $\mathcal{W}_3$]
\label{lemma_markov}
The $\mathcal{W}_{3}$ set includes the following groups:
\begin{equation}
 \Big\{ \mathbf{W}\big(u (a,b;t), w (a,b;t) \big) \Big| t\in[0,\infty) \Big\}
\end{equation}
where $b\in\mathbb{C},a\in\mathbb{R}$ denotes parameters characterizing the semigroup, and formulas for $u (a,b;t)$, $w (a,b;t)$ are given below:
\begin{itemize}
 \item for $a = 0$, $b = 0$
 \begin{equation}
  \left\{
  \begin{array}{rcl}
   u (a,b;t)
   &=&
   e^{-t}
   \\
   w (a,b;t) &=& 0
  \end{array}
  \right.
  ;
 \end{equation}
 \item if $\big|b\big|^2 - a^2 > 0$
 \begin{equation}
  \left\{
  \begin{array}{rcl}
   u (a,b;t)
   &=&
   \left(
    \cosh \big[\Delta t \big]
    + i \frac{a}{\Delta} \sinh \big[\Delta t \big]
   \right)
   e^{-t}
   \\
   w (a,b;t)
   &=&
   \frac{ b }{\Delta} \sinh \big[\Delta t \big]
   e^{-t}
  \end{array}
  \right.
  \qquad
  \text{where: }\Delta = \sqrt{ \big|b\big|^2 - a^2} ;
 \end{equation}

 \item if $a^2 - \big|b\big|^2 > 0$
 \begin{equation}
  \left\{
  \begin{array}{rcl}
   u (a,b;t)
   &=&
   \left(
    \cos \big[\Gamma t \big]
    + i \frac{a}{\Gamma} \sin \big[\Gamma t \big]
   \right)
   e^{-t}
   \\
   w (a,b;t)
   &=&
   \frac{ b }{\Gamma} \sin \big[\Gamma t \big]
   e^{-t}
  \end{array}
  \right.
  \qquad
  \text{where: }\Gamma = \sqrt{ a ^2 - \big|b\big|^2 } .
 \end{equation}

\end{itemize}
\end{lem}

Lemma \ref{lemma_markov} indicates that every curve representing a semigroup characterized by parameters $b\in\mathbb{C},a\in\mathbb{R}$, is tangent (for identity element) to a segment connecting identity with the point $(u,w) = \big(i a, b \big)$. This allows to conclude:

\begin{OurCorollary}[Non-dynamical $\arg{w}$ variable]
$\arg \big[ w(a,b;t) \big]$ remains constant for all matrices contained in a single semigroup.
 \begin{equation}
 \arg \big[ w(a,b;t) \big] = \arg \big[ b \big] \in \text{cons}(t).
 \end{equation}
 Furthermore, the values $u (a,b;t)$ and $\big| w (a,b;t) \big|$ depend only on
 $a$ and $|b|$, hence are independent of $\arg [ b ]$. As a consequence, the non-dynamic variable $\arg \big[ w(a,b;t) \big]$ separates from the rest of the problem.
\end{OurCorollary}

We conclude that the analysis of semigroups in $\mathcal{W}_3$ should be performed in reduced subsets of $\mathcal{W}_3$ with constant $\arg[w]$.
Such subsets represent tetrahedrons with height dependent on the value of $\arg[w]$.
An interactive applet, which generates semigroups as a function of input parameters $a$ and $|b|$
and draws their representations in 3D subsets of $\mathcal{W}_3$ with constant $\arg[w]$ may be found in the supplementary materials.

\section{Parameterization of the Birkhoff polytope of order $N=4$}
\label{sec:Parameterization4}

The parametrization procedure given in observation \ref{obs:parametryzacja} can be extended to $4\times4$ bistochastic matrices:
\begin{equation}
 \mathbf{W}(u,w_2,w_3,w_4,x)
 =
 \mathbf{B}_\star + x \mathbf{X} +
 + 2\operatorname{Re} \Big[ u \mathbf{D}_1 + w_2 \mathbf{D}_2 + w_3 \mathbf{D}_3  + w_4 \mathbf{D}_4 \Big]
 \in \mathcal{W}_4
\end{equation}

($u,w_2,w_3,w_4 \in \mathbb{C}$, $x\in\mathbb{R}$)
where:
\begin{align*}
 \mathbf{B}_\star
 &=
 \frac{1}{4}
 \left(
 \begin{smallmatrix}
  +1 & +1 & +1 & +1 \\
  +1 & +1 & +1 & +1 \\
  +1 & +1 & +1 & +1 \\
  +1 & +1 & +1 & +1 \\
 \end{smallmatrix}
 \right),
 &
 \mathbf{X}
 &=
 \frac{1}{4}
 \left(
 \begin{smallmatrix}
  +1 & -1 & +1 & -1 \\
  -1 & +1 & -1 & +1 \\
  +1 & -1 & +1 & -1 \\
  -1 & +1 & -1 & +1 \\
 \end{smallmatrix}
 \right),
 &&\\
 \mathbf{D}_1
 &=
 \frac{1}{4}
 \left(
 \begin{smallmatrix}
  +1 & +i & -1 & -i \\
  -i & +1 & +i & -1 \\
  -1 & -i & +1 & +i \\
  +i & -1 & -i & +1 \\
 \end{smallmatrix}
 \right),
 &
 \mathbf{D}_2
 &=
 \frac{1}{4}
 \left(
 \begin{smallmatrix}
  -i & +1 & +i & -1 \\
  +i & -1 & -i & +1 \\
  -i & +1 & +i & -1 \\
  +i & -1 & -i & +1 \\
 \end{smallmatrix}
 \right),
 &&\\
 \mathbf{D}_3
 &=
 \frac{1}{4}
 \left(
 \begin{smallmatrix}
  -1 & -i & +1 & +i \\
  -i & +1 & +i & -1 \\
  +1 & +i & -1 & -i \\
  +i & -1 & -i & +1 \\
 \end{smallmatrix}
 \right),
 &
 \mathbf{D}_4
 &=
 \frac{1}{4}
 \left(
 \begin{smallmatrix}
  +i & -i & +i & -i \\
  +1 & -1 & +1 & -1 \\
  -i & +i & -i & +i \\
  -1 & +1 & -1 & +1 \\
 \end{smallmatrix}
 \right) .
\end{align*}

This particular parametrization enables to introduce a lower-dimensional representation of $\mathcal{W}_4$ (cf. observation \ref{obs:parametryzacja})
\begin{equation}
 \mathbf{W}(u,w_2,w_3,w_4,x)
 \sim
   \begin{pmatrix}
   u   & \bar w_3 & w_4 \\
   w_3 & \bar u   & \bar w_4 \\
   w_2 & \bar w_2 & x \\
  \end{pmatrix}
  .
\end{equation}

Future studies will focus on multiplicative structure of the set of $4\times 4$ bistochastic matrices.

\section{Concluding remarks}

The present work demonstrates how the analysis of the dynamics of classical stochastic systems supports understanding of its quantum analogue.
As shown in section 2, the strategy for investigation of the set of bistochastic matrices can be understood as a preliminary step towards solution of the fully quantum problem in the three dimensional \textit{Hilbert space}. Next we focused on the description of Markovian dynamics in classical systems,
which were defined by appropriately chosen semigroups in the set of bistochastic matrices. The complex parametrization introduced in section 4 significantly simplified the analysis of the multiplicative structure of semigroups in $\mathcal{W}_3$. In consequence, multiplication of matrices in $\mathcal{W}_3$ was reduced to few operations on complex numbers (2 complex conjugates, four number multiplications and two additions).
This representation led to formulation of several theorems, indicating however that the topic is nothing but saturated.
A key point of the present study was extraction of an invariant coordinate associated with the Markov evolution in $\mathcal{W}_3$. The $4\times4$ case opens field for generalizations. Finally, a significant result from the quantum-information point of view, was the proof for no equality between the set of infinitely divisible time evolutions and the set of Markov evolutions.

\begin{acknowledgements}
We would like express our gratitude to K. \.Z{}yczkowski, for his encouragement to challenge this problem, and valuable remarks.
\end{acknowledgements}

\bibliographystyle{apsrev4-1}
\bibliography{Refs}

\end{document}